\documentclass[11pt,a4paper]{article}

\usepackage{epsfig}
\usepackage{amssymb}
\usepackage{graphicx}
\usepackage{color}
\usepackage{subfigure}

\usepackage{latexsym}

\begin{document}

\baselineskip 6mm
\renewcommand{\thefootnote}{\fnsymbol{footnote}}

\newcommand{\nc}{\newcommand}
\newcommand{\rnc}{\renewcommand}

%%%%%%%%%%%%%%%%%%%%%% Equation Numbering %%%%%%%%%%%%%%%%%%%%%%%
%\makeatletter \rnc{\theequation}{\thesection.\arabic{equation}}
%\@addtoreset{equation}{section} \makeatother

%%%%%%%%%%%%%%%%%%%%%%%%%%%%%%%%%%%%%%%%%%%%%%%%%%%%%%%%%%%%%%%%%
%                                                               %
%                NEW COMMANDS AND MACROS                        %
%                                                               %
%%%%%%%%%%%%%%%%%%%%%%%%%%%%%%%%%%%%%%%%%%%%%%%%%%%%%%%%%%%%%%%%%

\newcommand{\tcb}{\textcolor{blue}}
\newcommand{\tcr}{\textcolor{red}}
\newcommand{\tcg}{\textcolor{green}}

%%%%% Simplify some frequently used LaTeX commands %%%%%

\def\beq{\begin{equation}}
\def\eeq{\end{equation}}
\def\ba{\begin{array}}
\def\ea{\end{array}}
\def\bea{\begin{eqnarray}}
\def\eea{\end{eqnarray}}
\def\nn{\nonumber}

%%%%%%%%%%%%%%%%%%%%%%%%%%%%%%%%%%%%%
%  Journal 
%%%%%%%%%%%%%%%%%%%%%%%%%%%%%%%%%%%%

\def\CMP{Commun. Math. Phys.~}
\def\JHEP{JHEP~}
\def\Pre{Preprint}
\def\PRL{Phys. Rev. Lett.~}
\def\PR {Phys. Rev.~}
\def\CQG {Class. Quant. Grav.~}
\def\PL {Phys. Lett.~}
\def\NP {Nucl. Phys.~}

%%%%%%%%%%%%%%%%%%%%%%%%%%%%%%%%%%%%%%%%%%%%%%%%%%
%   Boldface Letters
%%%%%%%%%%%%%%%%%%%%%%%%%%%%%%%%%%%%%%%%%%%%%%%%%%
\def\G{\Gamma}

\def\S{{\bf S}}
\def\C{{\bf C}}
\def\Z{{\bf Z}}
\def\R{{\bf R}}
\def\N{{\bf N}}
\def\M{{\bf M}}
\def\P{{\bf P}}
\def\bm{{\bf m}}
\def\bn{{\bf n}}

\def\CA{{\cal A}}
\def\CB{{\cal B}}
\def\CC{{\cal C}}
\def\CD{{\cal D}}
\def\CE{{\cal E}}
\def\CF{{\cal F}}
\def\CH{{\cal H}}
\def\CM{{\cal M}}
\def\CG{{\cal G}}
\def\CI{{\cal I}}
\def\CJ{{\cal J}}
\def\CL{{\cal L}}
\def\CK{{\cal K}}
\def\CN{{\cal N}}
\def\CO{{\cal O}}
\def\CP{{\cal P}}
\def\CQ{{\cal Q}}
\def\CR{{\cal R}}
\def\CS{{\cal S}}
\def\CT{{\cal T}}
\def\CV{{\cal V}}
\def\CW{{\cal W}}
\def\CX{{\cal X}}
\def\CY{{\cal Y}}
\def\We{{W_{\mbox{eff}}}}

%%%%%%%%%%%%%%%%%%%%%%%%%%%%%%%%%%%%%%%%%%%%%%%%%%
%   Mathematical Symbols  AA
%%%%%%%%%%%%%%%%%%%%%%%%%%%%%%%%%%%%%%%%%%%%%%%%%%

\newcommand{\Lie}{\pounds}

\newcommand{\p}{\partial}
\newcommand{\bp}{\bar{\partial}}

\newcommand{\half}{\frac{1}{2}}

\newcommand{\bfalpha}{{\mbox{\boldmath $\alpha$}}}
\newcommand{\bfbeta}{{\mbox{\boldmath $\beta$}}}
\newcommand{\bfgamma}{{\mbox{\boldmath $\gamma$}}}
\newcommand{\bfmu}{{\mbox{\boldmath $\mu$}}}
\newcommand{\bfpi}{{\mbox{\boldmath $\pi$}}}
\newcommand{\bfvarpi}{{\mbox{\boldmath $\varpi$}}}
\newcommand{\bftau}{{\mbox{\boldmath $\tau$}}}
\newcommand{\bfeta}{{\mbox{\boldmath $\eta$}}}
\newcommand{\bfxi}{{\mbox{\boldmath $\xi$}}}
\newcommand{\bfkappa}{{\mbox{\boldmath $\kappa$}}}
\newcommand{\bfepsilon}{{\mbox{\boldmath $\epsilon$}}}
\newcommand{\bfTheta}{{\mbox{\boldmath $\Theta$}}}

\newcommand{\bz}{{\bar{z}}}

\newcommand{\dalpha}{\dot{\alpha}}
\newcommand{\dbeta}{\dot{\beta}}
\newcommand{\blambda}{\bar{\lambda}}
\newcommand{\btheta}{{\bar{\theta}}}
\newcommand{\bsigma}{{{\bar{\sigma}}}}
\newcommand{\bepsilon}{{\bar{\epsilon}}}
\newcommand{\bpsi}{{\bar{\psi}}}

%%%%%  Temporary notation %%%%

\def\ct{\cite}
\def\la{\label}
\def\eq#1{(\ref{#1})}

%%% Greek letters %%%

\def\a{\alpha}
\def\b{\beta}
\def\g{\gamma}
\def\G{\Gamma}
\def\d{\delta}
\def\D{\Delta}
\def\ep{\epsilon}
\def\e{\eta}
\def\ph{\phi}
\def\Ph{\Phi}
\def\ps{\psi}
\def\Ps{\Psi}
\def\k{\kappa}
\def\l{\lambda}
\def\L{\Lambda}
\def\m{\mu}
\def\n{\nu}
\def\th{\theta}
\def\Th{\Theta}
\def\r{\rho}
\def\s{\sigma}
\def\S{\Sigma}
\def\ta{\tau}
\def\o{\omega}
\def\O{\Omega}
\def\pr{\prime}

%%%%% Mathematical Symbols

\def\half{\frac{1}{2}}

\def\goto{\rightarrow}

\def\na{\nabla}
\def\grad{\nabla}
\def\curl{\nabla\times}
\def\div{\nabla\cdot}
\def\pa{\partial}

\def\bra{\left\langle}
\def\ket{\right\rangle}
\def\lb{\left[}
\def\lc{\left\{}
\def\ls{\left(}
\def\lp{\left.}
\def\rp{\right.}
\def\rb{\right]}
\def\rc{\right\}}
\def\rs{\right)}
\def\cl{\mathcal{l}}

\def\vac#1{\mid #1 \rangle}

%%%%  Special symbol
\def\td#1{\tilde{#1}}
\def\check{ \maltese {\bf Check!}}

%%%%% Roman pont in math

\def\Tr{{\rm Tr}\,}
\def\det{{\rm det}\,}

%%%%% Special format

\def\bc#1{\nnindent {\bf $\bullet$ #1} \\ }
\def\ch {$<Check!>$ }
\def\ss {\vspace{1.5cm}}

\begin{titlepage}

%---------------- preprint number ---------------
\hfill\parbox{5cm} { }

\hskip1cm

\vspace{10mm}

\begin{center}
%------------------------ title ------------------------
{\Large \bf Quasi-local  charges  and asymptotic symmetry generators}

%---------------- authors and addresses ----------------
\vskip 1. cm
  { Seungjoon Hyun\footnote{e-mail : sjhyun@yonsei.ac.kr}, Sang-A Park\footnote{e-mail : sangapark@yonsei.ac.kr},
  Sang-Heon Yi\footnote{e-mail : shyi@yonsei.ac.kr}
  }

\vskip 0.5cm

{\it Department of Physics, College of Science, Yonsei University, Seoul 120-749, Korea}
\end{center}

\thispagestyle{empty}

\vskip1.5cm

%----------------------- abstract ----------------------

\centerline{\bf ABSTRACT} \vskip 4mm

\vspace{1cm}
\noindent 
The quasi-local formulation of conserved charges through the off-shell approach is extended to cover the asymptotic symmetry generators. By introducing  identically conserved currents which are appropriate for asymptotic Killing vectors,  we show that the asymptotic symmetry generators can be understood  as  quasi-local charges. We also show that this construction is completely consistent with the on-shell method.
\vspace{2cm}
%PACS numbers:

%\today

\end{titlepage}

\renewcommand{\thefootnote}{\arabic{footnote}}
\setcounter{footnote}{0}

%%%%%%%%%%%%%%%%%%%%%%%%%%%%%%%%%%%%%%%%%%%%%%%%%%%%%%
\section{Introduction}
%%%%%%%%%%%%%%%%%%%%%%%%%%%%%%%%%%%%%%%%%%%%%%%%%%%%%
 Asymptotic symmetry generators in gravity have been known  to play  important roles in the AdS/CFT correspondence~\cite{Maldacena:1997re}, which should correspond to symmetry generators in the dual field theory. Especially to realize the infinite dimensional symmetry generators in the two-dimensional dual conformal field theory(CFT), the asymptotic symmetry generators  corresponding to asymptotic Killing vectors have been  constructed  in~\cite{Brown:1986nw} based on the Hamiltonian formalism. These generators are shown to form a Virasoro algebra with a non-vanishing central charge and the central charge has been used in various setups to reproduce the black hole entropy through the Cardy formula~\cite{Cardy:1986ie}. The successful outcomes include the explanation of the microscopic origin of black hole entropy~\cite{Strominger:1996sh}.  Since there have been much interests in the extension of the AdS/CFT correspondence to the space-time which is not  asymptotically AdS, the methodology to construct asymptotic symmetry generators in gravity is still important direction to be sought after.  One such direction has been the study on the asymptotic symmetry algebra in the context of  the Kerr/CFT correspondence~\cite{Guica:2008mu}.
 
 Among the asymptotic symmetry generators, there are generators or conserved charges which form a sub-algebra corresponding to the isometry group of the given geometry. Interestingly, it is not so straightforward  in gravity to identify even such generators or charges, and the method to obtain such charges has its own long history~\cite{Szabados:2004vb,Hollands:2005wt}. In particular, the concept of quasi-local conserved charges  is not yet firmly established and still causes some controversy.   To identify conserved charges at the asymptotic infinity, the traditional ADM formalism has been extended  to space-time with more generic asymptotic geometry by Abbott, Deser and Tekin(ADT)~\cite{Abbott:1981ff,Abbott:1982jh,Deser:2002rt,Deser:2002jk}.  Compared to other approaches, this ADT formalism has several merits. First of all, it is manifestly covariant and depends only on the equations of motion(EOM). Furthermore, it can be applied to a generic higher curvature theory of gravity. There is another covariant approach to conseved charges which is called the covariant phase space formalism~\cite{Lee:1990nz,Wald:1993nt,Iyer:1994ys,Wald:1999wa}. Contrary to the ADT formalism, this approach is based on the Lagrangian, not the equations of motion.  This covariant phase space formalism for conserved charges has been extended to include the asymptotic symmetry generators~\cite{Carlip:1999cy, Koga:2001vq}.  
 
 Yet another interesting covariant approach for asymptotic symmetry generators  was constructed by Barnich, Brandt and Comp\`{e}re(BBC) in~\cite{Barnich:2001jy,Barnich:2007bf,Compere:2007az}, which  is based on the, so-called, variational bi-complex. For the exact Killing vectors, the final expression of the symmetry generators in the BBC formalism turns out to be the same as the one from the covariant phase space formalism. In general, the expression of the asymptotic symmetry generators in this formalism  differs from  the one from the covariant phase space formalism. As a result, the central charge in the asymptotic symmetry algebra might be different. Still, both formalisms give the identical results for the set-ups in~\cite{Brown:1986nw,Hotta:2008yq,Compere:2008cv} and in~\cite{Guica:2008mu}. However, in the context of the Kerr/CFT correspondence in higher derivative theory of gravity,  the BBC formalism  gives the central charge, which, along with Cardy's formula,  is consistent with the Wald formula for the black hole entropy~\cite{Azeyanagi:2009wf}.

 Recently, the importance of the identically conserved or {\it off-shell} ADT current for a Killing vector is recognized and its applications are explored in Ref.~\cite{Kim:2013zha,Kim:2013cor}.  It was shown that one of  the interesting aspects of the off-shell construction is its intimate relationship with the quasi-local construction. In the end, the ADT formalism  for conserved charges was shown to give the same expression as those from the covariant phase space formalism, and thus also, those from the BBC formalism. This result naturally leads to the question on how to extend the ADT formulation to the case of asymptotic symmetry generators.  
 
 In this paper we would like to address this issue. We generalize  the off-shell ADT current and potential of the exact Killing vectors to those of the asymptotic Killing vectors.  In section 2, we construct the explicit form of the generalized off-shell ADT current. In section 3,  we give a natural way to obtain the generalized off-shell ADT potential from the Lagrangian. And then we compare our results with those in \cite{Barnich:2001jy,Koga:2001vq}. In the final section, we summarize our results and give some comments on the open issues and on the future direction.

%%%%%%%%%%%%%%%%%%%%%%%%%%%%%%%%%%%%%%%%%%%%%%%%%%%%%%
\section{Generalized off-shell conserved currents}
%%%%%%%%%%%%%%%%%%%%%%%%%%%%%%%%%%%%%%%%%%%%%%%%%%%%%%
In this section we introduce  the generalized off-shell ADT current in the spirit of the original construction of the ADT current~\cite{Abbott:1981ff,Abbott:1982jh,Deser:2002rt,Deser:2002jk}, which is based on the equations of motion(EOM). After reviewing the off-shell ADT current, we explain the necessity of its extension for asymptotic Killing vectors in the context of the AdS/CFT correspondence and then present its generic structure.  By using the integration by parts iteratively, we give a prescription to obtain the generalized off-shell ADT current unambiguously and present its explicit form depending only on the EOM and the linearized EOM expressions.
%%%%%%%%%%%%%%%%%%%%%%%%%%%%%%%%%%%%%%%%%%%%%%%%%%%%%%
\subsection{Off-shell currents}
%%%%%%%%%%%%%%%%%%%%%%%%%%%%%%%%%%%%%%%%%%%%%%%%%%%%%

We consider a generic theory of gravity with the action
\begin{equation} \label{action}
I[g] = \frac{1}{16\pi G}\int d^Dx~  \sqrt{-g}\, L(g)\,.
\end{equation}
For simplicity, we focus on the theory without any matter field in the following. The EOM of the metric  are given by $\CE^{\mu\nu}(g)=0$, whose Bianchi identity is $\nabla_\mu \CE^{\mu\nu}=0$.

In the geometry admitting a Killing vector $\xi^{\mu}$, the corresponding  
on-shell ADT current is introduced as  
\begin{equation} \label{on-shell}
\CJ^{\mu} (g\,;\, \xi,\delta g) = \delta \CE^{\mu\nu}\xi_{\nu}\,,
\end{equation} 
where $\delta$ denotes the variation with respect to the metric.
This on-shell current  can be shown to be conserved by using EOM, Bianchi identity and the Killing property of $\xi$. As a result, the anti-symmetric second rank tensor $Q^{\mu\nu}=Q^{[\mu\nu]}$, which is called the on-shell ADT potential, is introduced by $ \CJ^{\mu} = \nabla_{\nu}Q^{\mu\nu}$.
These on-shell current and potential are highly involved for a higher curvature/derivative theories of gravity, as their EOM are very complicated. Instead, the background independent ADT current and potential have been used for TMG~\cite{Bouchareb:2007yx} and new massive gravity~\cite{Nam:2010ub}. It was recognized that this background independent ADT current has not just  computational convenience in some specific theories but more profound meaning in a generic theory of gravity  as the off-shell extension of the on-shell ADT formulation for conserved charges~\cite{Kim:2013cor}.  

The off-shell ADT current for a Killing vector $\xi$ can be introduced as
\beq \label{offADT}
     {\cal J}^{\mu}_{ADT} (g\,;\, \xi,\delta g)  \equiv  \delta \CE^{\mu\nu}\xi_{\nu} + \frac{1}{2}g^{\alpha\beta}\delta g_{\alpha\beta}\, \CE^{\mu\nu}\xi_{\nu}  + \CE^{\mu\nu}\delta g_{\nu\rho}\, \xi^{\rho}- \frac{1}{2}\xi^{\mu}\CE^{\alpha\beta}\delta g_{\alpha\beta} \,. \label{ADT}
\eeq
This off-shell current can be shown to be identically conserved by using the Bianchi identity and the Killing property of the vector $\xi$.  One may note that this current reduces to the on-shell current in Eq.~(\ref{on-shell}) after using the EOM, $\CE^{\mu\nu}(g)=0$. This identical conservation property allows us to introduce the off-shell ADT potential $Q^{\mu\nu}_{ADT} $ for a Killing vector $\xi$ as  
\beq \label{ADTQ} 
\CJ_{ADT}^{\mu} (g\,;\, \xi,\delta g) = \nabla _{\nu}Q^{\mu\nu}_{ADT} (g\,;\, \xi,\delta g) \,.  
 \eeq
By using the above off-shell ADT potential and the one-parameter path in the solution space~\cite{Kim:2013cor, Barnich:2003xg},  one can introduce quasi-local conserved charges for the Killing vector $\xi$ as
\beq  Q(\xi) = \frac{1}{8\pi G}\int^{1}_{0}ds \int_{\CB} d^{D-2} x_{\mu\nu} \sqrt{-g}\, Q^{\mu\nu}_{ADT}(g; \xi | s)\,,
\label{ADTcharge}
\eeq
where $\CB$ may be taken in the interior region not just at the asymptotic infinity of the space-time. One may note that quasi-local charges are computed, at the end,  on the on-shell value since we have used a one-parameter path in the solution space.  As was shown in~\cite{Kim:2013zha,Kim:2013cor},  these quasi-local charges for  the Killing vector associated with a Killing horizon reproduce the Wald's entropy formula for black holes  and those computed at the asymptotic infinity coincide with the original ADT charges.

In view of AdS/CFT correspondence, it is desirable to extend this formulation to the case with asymptotic symmetry generators,  which may be realized as the conserved charges for the asymptotic Killing vectors.  Recall that the off-shell ADT current, which is introduced in Eq.(\ref{offADT}), is conserved  for an exact Killing vector but not for an asymptotic Killing vector. Rather, for an asymptotic Killing vector $\zeta$, it satisfies  
\begin{eqnarray}  
\p_{\mu}(\sqrt{-g}\CJ^{\mu}_{ADT} ) &=&  \delta (\sqrt{-g} \CE^{\mu\nu})\nabla_{(\mu}\zeta_{\nu)} + \frac{1}{2}\sqrt{-g}\CE^{\mu\nu}\Lie_{\zeta}\delta g_{\mu\nu}   \nn \\ 
&& \qquad   - \frac{1}{2}\sqrt{-g}\Big[\nabla_{\mu}\zeta^{\mu}\CE^{\alpha\beta}\delta g_{\alpha\beta} + \Lie_{\zeta}(\CE^{\alpha\beta}\delta g_{\alpha\beta} ) \Big]\nn \\
&=&
 \frac{1}{2}\Big[\delta (\sqrt{-g} \CE^{\mu\nu})\Lie_{\zeta}g_{\mu\nu}  - \Lie_{\zeta}( \sqrt{-g}\CE^{\alpha\beta})\, \delta g_{\alpha\beta} \Big]\,,\label{div}
\end{eqnarray}
which shows us that the off-shell ADT current $\CJ^{\mu}_{ADT}$ is not conserved  for asymptotic Killing vectors  and needs to be extended.

In this paper, we would like to generalize the above off-shell ADT current for a Killing vector $\xi$ to the off-shell current for an asymptotic Killing vector $\zeta$. 
By noting that this generalized off-shell ADT current should depend linearly on the vector $\zeta$ and reduce to the off-shell ADT current ${\CJ}^{\mu}_{ADT}$ when $\zeta$ is taken as a Killing vector, one may take the generalized off-shell ADT current ${\bf J}^{\mu}_{ADT}$, without loss of generality,  in the form of 
\bea \label{}
{\bf J}^{\mu}_{ADT}(\zeta) &=& \CM^{\mu\nu}\zeta_{\nu} + \CM^{\mu \alpha\beta}\Lie_{\zeta}g_{\alpha\beta} +  \nabla_{\nu_{1}}(\CM^{\mu \nu_{1} \alpha\beta} \Lie_{\zeta}g_{\alpha\beta} )    +  \nabla_{\nu_1}\nabla_{\nu_2}( \CM^{\mu \nu_1 \nu_2 \alpha\beta} \Lie_{\zeta}g_{\alpha\beta} )  ~~~~~ \\
&& + ~\nabla_{\nu_1}\nabla_{\nu_2} \nabla_{\nu_3}( \CM^{\mu \nu_1 \nu_2 \nu_3 \alpha\beta}\Lie_{\zeta}g_{\alpha\beta}) + \cdots +\nabla_{\nu_1}\cdots \nabla_{\nu_n}(\CM^{\mu \nu_1 \cdots \nu_n}\Lie_{\zeta}g_{\alpha\beta})  \,, \nn
\eea
where  $\Lie_{\zeta}$ denotes the Lie derivative along $\zeta$ direction. 
$\CM^{\mu \nu_1\cdots \nu_k \alpha\beta}$'s are taken such that they satisfy the following  properties
\beq \label{}
\CM^{\mu\nu}\zeta_{\nu} = \CJ^{\mu}_{ADT}(\zeta)\,, \qquad 
\CM^{\mu \nu_1\cdots \nu_k \alpha\beta} = \CM^{\mu (\nu_1\cdots \nu_k) \alpha\beta}\,,
\eeq
where the round parenthesis denotes the total symmmetrization with a normalization factor $1/k!$. 
One can always take these forms of $\CM$'s and ${\bf J}^{\m}_{ADT}$ by using the fact that any commutator of $\nabla$'s can be replaced by the Riemann tensor. The identical conservation of ${\bf J}^{\m}_{ADT}$ leads to severe constraints on the form of $\CM$'s. Rather than solving these constraints directly, we will take a different methodology and propose a way to obtain the unambiguous form of the generalized off-shell ADT current in the following section.

%%%
%%%%%%%%%%%%%%%%%%%%%%%%%
\subsection{Construction}
%%%%%%%%%%%%%%%%%%%%%%%%%

We introduce the generalized off-shell ADT current for an asymptotic Killing vector $\zeta$ as 
\begin{equation} \label{genoffADT}
{\bf J}^{\mu}_{ADT} (g\,;\, \zeta, \delta g) =  \CJ^{\mu}_{ADT} (g\,;\,  \zeta, \delta g) + \CJ^{\mu}_{\Delta}(g\,;\, \Lie_{\zeta}g, \delta g)  \,.
\end{equation}
The second term $\CJ^{\mu}_{\Delta}$  is introduced to preserve the off-shell conservation of the ADT current for an asymptotic Killing vector $\zeta$, such that  its divergence cancels the right-hand side of Eq.~(\ref{div}).  To obtain $\CJ^{\mu}_{\Delta}$ explicitly,
let us consider an $n$-th order derivative theory of gravity. Generically the linearized EOM expression $\CE^{\mu\nu}$ can be written as
\bea \label{deltaE}
\delta (\sqrt{-g}\CE^{\mu\nu}) 
&=& \sqrt{-g} \Big[ f^{\mu\nu\alpha\beta}\delta g_{\alpha\beta} +\sum_{k=1}^{n}f^{\mu\nu\alpha\beta \,|\, \rho_1\cdots \rho_k}\nabla_{(\rho_1}\cdots \nabla_{\rho_k)}\delta g_{\alpha\beta}  \Big]\,, 
\eea
where the  coefficient functions $f$'s satisfy 
\begin{equation} \label{}
f^{\mu\nu\alpha\beta\,|\, \rho_1 \cdots \rho_k}(g) =f^{\mu\nu\alpha\beta\,|\, (\rho_1 \cdots \rho_k)}(g)=f^{(\mu\nu)\alpha\beta\,|\, \rho_1 \cdots \rho_k}(g) = f^{\mu\nu(\alpha\beta)\,|\, \rho_1 \cdots \rho_k}(g)  \,.
\end{equation}
The above form of the linearized EOM may be regarded as a generic derivative expansion since the commutators of the covariant derivatives can always be replaced by Riemann tensors.  Because of the symmetrization, the order of covariant derivatives does not matter. {}From now on, we will always order the covariant derivatives in the increasing $\rho_i$-numbering just for the convenience. 

As noted above, the additional current $\CJ^{\mu}_{\Delta}$ is designed to satisfy  the following relation
\begin{equation} \label{Jdelta}
\delta (\sqrt{-g} \CE^{\mu\nu})\Lie_{\zeta}g_{\mu\nu}  - \Lie_{\zeta}( \sqrt{-g}\CE^{\alpha\beta})\, \delta g_{\alpha\beta} = -\p_{\mu}(2\sqrt{-g}\CJ^{\mu}_{\Delta})\,.
\end{equation}
By using the expression in (\ref{deltaE}), the first term in the left-hand side can be written generically in the form as 
\bea \label{Intbypart}
\delta (\sqrt{-g}\CE^{\mu\nu})\, \Lie_{\zeta}g_{\mu\nu} &=& \sqrt{-g} \Big[ f^{\mu\nu\alpha\beta}\delta g_{\alpha\beta} +\sum_{k=1}^{n}f^{\mu\nu\alpha\beta \,|\, \rho_1\cdots \rho_k}\nabla_{(\rho_1}\cdots \nabla_{\rho_k)}\delta g_{\alpha\beta}  \Big]\Lie_{\zeta}g_{\mu\nu}  \nonumber \\
&=&
\sqrt{-g}\CF^{\alpha\beta}\delta g_{\alpha\beta} +\p_{\r}(\sqrt{-g}{\cal H}^{\r})\,,
\eea
where we performed the integration by parts iteratively to arrive at the second equality. 
The explicit forms of $\CF^{\alpha\beta}$ and ${\cal H}^{\r}$ are given by 
\footnote{As mentioned earlier, the covariant derivatives are ordered in the increasing $\r_i$ indices. Furthermore, in order to express the formula compactly, we also adopt the summation convention such that there is no covariant derivative if the left hand covariant derivative numbering is greater than the right hand one, i.e. 
$\nabla_{\r_{k+1}}\nabla_{\r_{k}}\d g_{\a\b}$ denotes $\d g_{\a\b}$ under the $k$ summation.}
\bea \label{ours}
\CF^{\alpha\beta}  (\Lie_{\zeta}g) &=&   f^{\mu\nu\alpha\beta}\, \Lie_{\zeta}g_{\mu\nu} + \sum_{k=1}^{n}(-1)^k\nabla_{\rho_1}\cdots\nabla_{\rho_k}(f^{\mu\nu\alpha\beta\,|\, \rho_1\rho_2\cdots\rho_k}\, \Lie_{\zeta}g_{\mu\nu})\,,
\nn \\
{\cal H}^{\rho}\, (\Lie_{\zeta}g,\, \delta g) &=&  f^{\mu\nu\alpha\beta \,|\, \rho}\Lie_{\zeta}g_{\mu\nu}~ \delta g_{\alpha\beta} + \sum_{k=2}^{n} f^{\mu\nu\alpha\beta\,|\, \rho\rho_2\cdots\rho_k}\Lie_{\zeta}g_{\mu\nu}\nabla_{\rho_2}\cdots\nabla_{\rho_k}\delta g_{\alpha\beta}~~~~~  \\ && \quad +  \sum_{k=2}^{n} \sum_{l=2}^{k}(-1)^{\ell-1}\nabla_{\rho_2}\cdots\nabla_{\rho_l}( f^{\mu\nu\alpha\beta \,|\, \rho \rho_2\cdots \rho_{k}}\, \Lie_{\zeta}g_{\mu\nu})~ \nabla_{\rho_{l+1}}\cdots \nabla_{\rho_k}\delta g_{\alpha\beta}   \,. \qquad  \nn
\eea
As shown in the appendix A, it turns out that
\begin{equation} \label{Ida}
\sqrt{-g}\CF^{\mu\nu} (\Lie_{\zeta} g) = \Lie_{\zeta} (\sqrt{-g}\CE^{\mu\nu})\,.
\end{equation}
By combining the equations,  (\ref{Jdelta}), (\ref{Intbypart}) and  the relation  (\ref{Ida}), we can naturally identify the additional current $\CJ^{\r}_{\Delta} $ as 
\begin{equation}\label{sang}
\CJ^{\r}_{\Delta} (\Lie_{\zeta}g,\, \delta g) \equiv -\frac{1}{2} {\cal H}^\r(\Lie_{\zeta}g,\, \delta g) \,.
\end{equation}

In general, the identical conservation condition alone does not give us the unique expression of the additional current $\CJ^{\r}_{\Delta}$.
In the above, we have fixed this ambiguity by declaring that $\CJ_{\Delta}^{\r}$ is constructed only by $f$ functions as  in Eq.s~(\ref{ours}) and (\ref{sang}). In other words,
by performing the integration by parts successively, $\CJ^{\r}_{\D}$ can be rewritten  in the following form of
\begin{equation} \label{Criterion}
\CJ^{\r}_{\Delta}\, (\Lie_{\zeta}g, \delta g) = \delta g_{\alpha\beta}\,\CV^{\alpha\beta\,|\, \rho}(\Lie_{\zeta}g) + \nabla_{\rho'} {\bf S}^{\rho\rho'}(\Lie_{\zeta}g,\delta g)\,, 
\end{equation}
where ${\bf S}^{\rho\rho'}$ should be a symmetric tensor over indices $\rho$ and $\rho'$. This is the criterion for  fixing the ambiguity in our approach. 

As shown in appendix {\bf B}, one can obtain more useful expression  of $\CF^{\alpha\beta}$ in Eq. (\ref{ours}) as
\begin{eqnarray}  
\CF^{\alpha\beta}
&=&  \sum_{p=0}^{n}\sum_{k=p}^{n}(-1)^k{ k \choose p}\nabla_{\rho_{p+1}}\cdots \nabla_{\rho_k} f^{\mu\nu\alpha\beta\,|\, \rho_1\cdots \rho_k}~ \nabla_{\rho_1}\cdots \nabla_{\rho_{p}}\Lie_{\zeta}g_{\mu\nu}\,,   \label{Fexp} \end{eqnarray}
and
 $\CJ^{\r}_{\Delta}$ in Eq. (\ref{sang}) can be written in the form of 
\begin{eqnarray}  
\CJ^{\rho}_{\Delta}(\Lie_{\zeta}g, \delta g) &=&\frac{1}{2} \sum_{p=1}^{n}\sum_{k=p}^{n}\sum_{q=0}^{\big[\frac{p-1}{2}\big]}(-1)^{p+k+q+1} {k-p +q \choose q}~   \nabla_{\rho_{p+1}}\cdots \nabla_{\rho_k} f^{\mu\nu\alpha\beta\,|\, \rho\rho_2\cdots\rho_k}     \label{CJexpF}\\ 
&&  \quad   \times \Big(1- \frac{1}{2}\,\delta_{2q+1, p}\Big)  \Big(\nabla_{\rho_{p-q+1}}\cdots\nabla_{\rho_p}\Lie_{\zeta}g_{\mu\nu}\, \nabla_{\rho_2}\cdots \nabla_{\rho_{p-q}}\delta g_{\alpha\beta}  \nn \\
&& \qquad \qquad \qquad \qquad \qquad \qquad  -~ \nabla_{\rho_2}\cdots \nabla_{\rho_{p-q}}\Lie_{\zeta} g_{\alpha\beta}  \nabla_{\rho_{p-q+1}}\cdots\nabla_{\rho_p}\delta g_{\mu\nu}\, \Big)\,. \nn 
\end{eqnarray}
This expression of $\CJ^{\rho}_{\Delta}(\Lie_{\zeta}g, \delta g)$ is one of our main results. By construction,  the additional  current $\CJ^{\r}_{\Delta}$ depends only on the EOM.  It  is manifestly covariant and conserved when $g$ satisfies the EOM and $\delta g$ does the linearized EOM. Furthermore, it turns out to be a  symplectic current as will be shown in the next section.

Now, it is straightforward to see that this current is antisymmetric about its arguments:
\begin{equation} \label{}
\CJ^{\rho}_{\Delta}(\Lie_{\zeta}g, \delta g) =  -\CJ^{\rho}_{\Delta}(\delta g, \Lie_{\zeta}g)\,.
\end{equation}
One might say that the demanded condition of $\CJ^{\mu}_{\Delta}$ in Eq.~(\ref{Jdelta}) implies the above antisymmetric property. We would like to emphasize that the demanded condition does not warrant the above antisymmetric property  since it does not  fix the ambiguity which allows the addition of  total derivative terms. In contrast, we constructed the explicit form of $\CJ^{\rho}_{\Delta}(\Lie_{\zeta}g, \delta g)$ in Eq.~(\ref{CJexpF}) with manifestly antisymmetric property without any ambiguity. 

A few leading order terms of $\CJ^{\rho}_{\Delta}$ are given explicitly as 
\begin{eqnarray}  
\!\!\!\!\!\!\!\!\!\!\!\! && \CJ^{\rho}_{\Delta} (\Lie_{\zeta}g, \delta g)  ~ = \nn \\    
 &&~~~~~  \frac{1}{2}
\sum^{n}_{k=1}(-1)^k\nabla_{\rho_2}\cdots \nabla_{\rho_k} f^{\mu\nu\alpha\beta\,|\, \rho\rho_2\cdots \rho_k}~ \frac{1}{2} \Big[\Lie_{\zeta}g_{\mu\nu}\, \delta g_{\alpha\beta} - \Lie_{\zeta}g_{\a\b}\, \delta g_{\m\n} \Big]\nn 
\\
&&~~~~ -\frac{1}{2} \sum_{k=2}^{n}(-1)^k\nabla_{\rho_3}\cdots \nabla_{\rho_k} f^{\mu\nu\alpha\beta\,|\, \rho\rho_2\cdots \rho_k}~ \Big[\Lie_{\zeta}g_{\mu\nu}\, \nabla_{\rho_2} \delta g_{\alpha\beta} - \nabla_{\rho_2}\Lie_{\zeta}g_{\mu\nu}\,  \delta g_{\alpha\beta} \Big]    \nn \\
 &&~~~~ + \frac{1}{2} \sum_{k=3}^{n}(-1)^k\nabla_{\rho_4}\cdots \nabla_{\rho_k} f^{\mu\nu\alpha\beta\,|\, \rho\rho_2\cdots \rho_k}~ \Big[\Big(\Lie_{\zeta}g_{\mu\nu}\, \nabla_{\rho_2}\nabla_{\rho_3} \delta g_{\alpha\beta}  - \nabla_{\rho_2}\nabla_{\rho_3}\Lie_{\zeta}g_{\mu\nu}\,  \delta g_{\alpha\beta} \Big)\nn \\
&&~~~~ \qquad \qquad \qquad  \qquad \qquad  \qquad \qquad   - \frac{1}{2}(k-2)\Big(\nabla_{\rho_2}\Lie_{\zeta} g_{\mu\nu} \nabla_{\rho_3}\delta g_{\alpha\beta} -  \nabla_{\rho_3}\Lie_{\zeta} g_{\mu\nu} \nabla_{\rho_2}\delta g_{\alpha\beta} \Big)\Big]   \nn \\
&&~~~~  + \cdots\,.  
\end{eqnarray}
For example,  in the case of Einstein gravity, the  expression of $\CJ^{\rho}_{\Delta} (\Lie_{\zeta}g, \delta g) $ from our formula is given by
\begin{eqnarray}  
\CJ^{\rho}_{\Delta} (\Lie_{\zeta}g, \delta g)   =  -\frac{1}{2} f^{\mu\nu\alpha\beta\, |\, \rho \rho_2}\Big[\Lie_{\zeta}g_{\mu\nu}\, \nabla_{\rho_2} \delta g_{\alpha\beta} - \nabla_{\rho_2}\Lie_{\zeta}g_{\mu\nu}\,  \delta g_{\alpha\beta} \Big]\,,  
\end{eqnarray}
where
\begin{eqnarray}
 f^{\mu\nu\alpha\beta\, |\, \rho_1 \rho_2}&=&\frac{1}{2} \Big[g^{\mu\nu}g^{\alpha\beta}g^{\rho_1\rho_2}  -  g^{\mu\nu}g^{\alpha(\rho_1}g^{\rho_2)\beta} - g^{\alpha\beta}g^{\mu(\rho_1}g^{\rho_2)\nu} \nn \\
&&  \qquad - g^{\alpha(\mu}g^{\nu)\beta}g^{\rho_1\rho_2} + g^{\rho_1(\mu}g^{\nu)(\alpha}g^{\beta)\rho_2} + g^{\rho_2(\mu}g^{\nu)(\alpha}g^{\beta)\rho_1} \Big]\,. \nn 
\end{eqnarray}
It is interesting to note that the expression of $\CJ^{\mu}_{\Delta}$  is exactly half of the expression of the so-called invariant symplectic current $W^{\mu}$  at (E.15) in~\cite{Barnich:2007bf} up to the sign convention. 

Since we have constructed $\CJ^{\mu}_{\Delta}$ explicitly, we can obtain the generalized off-shell ADT current ${\bf J}^{\mu}_{ADT}$. Furthermore, by Poincar\'{e} lemma,  one can introduce the generalized off-shell ADT potential ${\bf Q}^{\mu\nu}_{ADT}$ as 
\begin{eqnarray}  \label{genoffADTpot}
\sqrt{-g}\, {\bf J}^{\mu}_{ADT} (g\,;\, \xi, \delta g)  &\equiv&  \p_{\nu}\Big(\sqrt{-g}\, {\bf Q}^{\mu\nu}_{ADT} (g\,;\,\xi, \delta g)\, \Big)   \,.
\end{eqnarray}
In the following section, we explore the connection of our construction, which depends only on the EOM, to the covariant phase space approach, and then show how to obtain the above generalized off-shell ADT potential from the Lagrangian. 

%%%%%%%%%%%%%%%%%%%%%%%%%
\section{Off-shell potential and asymptotic symmetry generators}
%%%%%%%%%%%%%%%%%%%%%%%%%
In this section, we compare our construction to the covariant phase space approach~\cite{Lee:1990nz,Wald:1993nt,Iyer:1994ys} and indicate the difference explicitly. Then we present the method  to obtain the generalized off-shell ADT potential from the given Lagrangian. As a specific example, we consider a generic higher curvature theory of gravity. We also give some comments on the relation of our construction to a mathematical construction based on the so-called variational bi-complex~\cite{Barnich:2001jy,Compere:2007az}. 

%%%%%%%%%%%%%%%%%%%%%%%%%
\subsection{Comparison with the covariant phase space}
%%%%%%%%%%%%%%%%%%%%%%%%%
It has been known that there are several ways to construct conserved charges for an asymptotic Killing vectors~\cite{Brown:1986nw,Brown:1992br,Barnich:2001jy,Barnich:2003xg,Hotta:2008yq}.
Though we have constructed quasi-local conserved charges through the additional current $\CJ_{\D}^{\m}$ explicitly, 
it is useful to find the connection of our construction with the covariant phase space method which is based on the Lagrangian. For simplicity, we will focus on a covariant theory of gravity. It is straightforward to include gravitational Chern-Simons terms, which will be omitted.

The variation of the action (\ref{action}) with respect to  $g^{\mu\nu}$  is  taken in the form of
\beq
\delta I[g] =   \frac{1}{16\pi G}\int d^Dx~  \delta (\sqrt{-g}\CL) =  \frac{1}{16\pi G}\int d^{D}x  \Big[  \sqrt{-g} \CE_{\mu\nu}\delta g^{\mu\nu}+\p_{\mu}\Theta^{\mu}(\delta g) \Big]\,, \label{genVar}
\label{firstvary}
\eeq
where  $\Theta^{\m}$ denotes a surface term. By using the surface term $\Theta^{\mu}$, one can introduce the so-called symplectic current $\omega^{\m}$ as
\beq
\omega^{\m}(g\,;\, \delta_1g, \delta_2 g)  \equiv  \delta_1 \Theta^{\m}(g\,;\, \delta_2 g) - \delta_2 \Theta^{\m}(g\,;\, \delta_1 g)\,,
\label{symplectic}
\eeq
which is symplectic in the sense that it satisfies~\cite{Lee:1990nz}
\[
\delta_1\omega^{\m}(g\,;\, \delta_2g, \delta_3 g) + \delta_2\omega^{\m}(g\,;\, \delta_3g, \delta_1 g) + \delta_3\omega^{\m}(g\,;\, \delta_1g, \delta_2 g) =0\,.
\]
By applying another variation to Eq.~(\ref{genVar}) and using $\delta_1\delta_2I[g] = \delta_2\delta_1I[g]$ with $\delta_1\delta_2g_{\mu\nu}=\delta_2\delta_1 g_{\mu\nu}$, one can show that this symplectic current  satisfies the following relation 
\begin{equation} \label{}
\p_{\mu}\omega^{\mu} (g\,;\, \delta_1g, \delta_2 g) = \delta_1(\sqrt{-g}\CE^{\mu\nu})\delta_2g_{\mu\nu} -  \delta_2(\sqrt{-g}\CE^{\mu\nu})\delta_1g_{\mu\nu}\,.
\end{equation}
This relation shows us that $\omega^{\mu}$ is a  conserved current when $g$ and $\delta g$ satisfy the EOM and the linearized EOM, respectively.  Furthermore,  one may notice  that  the symplectic current $\omega^{\m}(g\,;\, \Lie g, \delta g) $ satisfies the same divergence  relation with our additional current $\CJ_{\D}^{\m}(\Lie g, \delta g)$ in Eq.~(\ref{Jdelta}). This means that the difference between them should be, at most, a total derivative as
\beq \label{Adef}
 2\sqrt{-g}\CJ^{\mu}_{\Delta} (g\,;\, \Lie g, \delta g) = \omega^{\mu} (g\,;\, \Lie g, \delta g)  +\p_{\nu}\Big(\sqrt{-g}{\bf A}^{\mu\nu}(g\,;\, \Lie g, \delta g) \Big)\,,
\eeq
where ${\bf A}^{\mu\nu}$ denotes a certain antisymmetric second rank tensor determined by the given Lagrangian. 
Note that this relation tells us  that $2\sqrt{-g}\CJ^{\mu}_{\Delta}$ is symplectic up to a total derivative. 
And then, the generalized off-shell ADT current ${\bf J}^{\mu}_{ADT}$ can  be written in terms of $\omega^{\mu}$ as
\begin{equation} \label{}
 \sqrt{-g}\, {\bf J}^{\mu}_{ADT} (g, \xi, \delta g)  = \p_{\nu}(\sqrt{-g}{\bf Q}^{\mu\nu}_{ADT}) =  \sqrt{-g} \CJ^{\mu}_{ADT} +\frac{1}{2} \omega^{\mu}  + \p_{\nu}\Big(\frac{1}{2}\sqrt{-g}{\bf A}^{\mu\nu}\Big) \,.
\end{equation}

In order to see  the relation between our construction and the covariant phase space approach, one may note that $\CJ^{\mu}_{ADT}$ vanishes whenever $g$ and $\delta g$ satisfy the EOM and the linearized EOM, respectively.  Under this condition, the generalized off-shell ADT current ${\bf J}^{\mu}_{ADT}$ reduces to the symplectic current $\omega^{\mu}$ up to a total derivative. Hence, the  on-shell difference between our construction and the covariant phase space formalism resides in the total derivative term ${\bf A}^{\mu\nu}$.  Now, we give a recipe to determine ${\bf A}^{\mu\nu}$ from the surface term $\Theta^{\m}$. By recalling  the  form of $\CJ^{\mu}_{\Delta}$ obtained in Eq.~(\ref{Criterion}), the definition of $\omega^{\m}$ given in Eq.~(\ref{symplectic}) and the relation between them in Eq.~(\ref{Adef}), one can set
\begin{equation} \label{ATheta}
 \delta \Theta^{\mu}(\Lie_{\zeta}g) =  \Lie_{\zeta}{\Theta}^{\mu}(\delta g)    +  \delta g_{\alpha\beta}\Big[ -2\sqrt{-g}\, \CV^{\alpha\beta\,|\, \mu}(\Lie_{\zeta}g) \Big]   +  \sqrt{-g} \nabla_{\nu} \Big( {\bf A}^{\mu\nu}(\Lie_{\zeta}g,\delta g) - 2{\bf S}^{\mu\nu}(\Lie_{\zeta}g,\delta g) \Big)\,,
\end{equation}
where one may notice that the total derivative part is composed of two pieces, symmetric ${\bf S}^{\mu\nu}$ and antisymmetric ${\bf A}^{\mu\nu}$. 
This relation gives us a definite way to obtain ${\bf A}^{\mu\nu}$ from the variation of  the surface term $\Theta^{\m}$.  
Schematically, the $\Theta$ term is given by  $\Theta(\delta g)  \sim \nabla\cdots \nabla \delta g$. Therefore, one can see that 
\[   
\delta \Theta(\Lie g) ~ \sim ~ \nabla\cdots \delta \nabla \Lie g + \nabla\cdots \nabla \delta \Lie g ~  \sim ~ \nabla\cdots \delta \nabla \Lie g + \nabla\cdots \nabla \Lie \delta  g  \,,
\]
where we have used $\Lie \delta g = \delta \Lie g$. To relate this form to the expression of $\Lie \Theta(\delta g)$, it is useful to note the following identity for arbitrary tensor $T_{\mu_1\cdots \mu_k}$:
\begin{equation} \label{}
[\nabla_{\rho}, \Lie_{\zeta}] T_{\mu_1\cdots \mu_k} = \frac{1}{2}T^{~~~~~~~~~ \nu}_{\mu_1\cdots \mu_{i-1}~~ \mu_{i+1}\cdots \mu_k} \Big(\nabla_{\rho}\Lie_{\zeta}g_{\mu_i \nu} + \nabla_{\mu_i}\Lie_{\zeta}g_{\rho \nu} - \nabla_{\nu}\Lie_{\zeta}g_{\rho \mu_i} \Big)\,.
\end{equation}
By performing the integration by parts on the variation of the covariant derivatives $\delta \nabla$, one can extract ${\bf A}^{\mu\nu}$ from the anti-symmetric part, over $\mu\nu$-indices, inside  the total derivatives.  The explicit examples are given in the next section.

Though we have constructed ${\bf J}^{\mu}_{ADT}$ and ${\bf Q}^{\mu\nu}_{ADT}$ from the EOM expression, it is much better to obtain the generalized off-shell ADT ${\bf Q}^{\mu\nu}_{ADT}$ from the given Lagrangian. By doing this, one can compare more clearly our construction to the covariant phase space approach or the BBC formalism. To this purpose,
it is useful to recall  the Noether current  and potential.
The off-shell Noether current $J^{\m}$ under the general diffeomorphism $\zeta$  can be introduced as
\beq J^{\mu}(\zeta) = 2\sqrt{-g}\, \CE^{\mu\nu}\zeta_{\nu} +  \sqrt{-g}\, \zeta^{\mu} L    - \Theta^{\mu}(\Lie_{\zeta}g) \,. \label{OffNoether} \eeq
This current can be shown to be identically conserved and thus the associated potential $K^{\m\n}$  can be defined by
\beq J^{\mu}(\zeta)  \equiv \p_{\nu}K^{\mu\nu} (\zeta)\,.  \eeq

Now we derive the relation among the generalized off-shell ADT current, the $K^{\mu\nu}$ term and  the $\Theta^{\mu}$ term for the asymptotic Killing vectors, following the steps  in the case of  exact Killing vectors~\cite{Kim:2013zha}.
By taking a generic metric variation of the off-shell Noether current, one can obtain
\begin{eqnarray}  
\p_{\nu}\delta K^{\mu\nu}  &=& 2\delta (\sqrt{-g}\, \CE^{\mu\nu}\zeta_{\nu}) +  \delta(\sqrt{-g}\, \zeta^{\mu} L)    - \delta \Theta^{\mu}(\zeta)  \nn \\
&=& 2\sqrt{-g}\CJ^{\mu}_{ADT} + \Lie_{\zeta} \Theta^{\mu}(\delta g) - \delta \Theta^{\mu}(\zeta) + \p_{\nu}\Big(2\zeta^{[\mu}\Theta^{\nu]}(\delta g) \Big)\,, \nn 
\end{eqnarray}
where we have used the generic variation of the Lagrangian in Eq.~(\ref{genVar}) and the definition of Lie derivative on the $\Theta$ term as
\[   
\Lie_{\zeta} \Theta^{\mu} = \zeta^{\nu}\p_{\nu}\Theta^{\mu} - \Theta^{\nu}\p_{\nu}\zeta^{\mu} + \Theta^{\mu} \p_{\nu}\zeta^{\nu}\,.
\]
This leads to the identity for the general diffeomorphism $\zeta$:
\beq \label{}
 2\sqrt{-g} \CJ^{\mu}_{ADT} + \omega^{\mu}(g\, ;\, \Lie_{\zeta} g, \delta g) = \p_{\nu}\Big( \delta K^{\mu\nu} (\zeta)- 2 \zeta^{[\mu}\Theta^{\nu]}(g;\, \delta g) \Big)\,.
\eeq
As a result,  our final expression of the generalized off-shell ADT potential ${\bf Q}_{ADT}^{\m\n}$ is given by
\beq 2\sqrt{-g}\, {\bf Q}^{\mu\nu}_{ADT}(g\,;\, \zeta, \delta g) = \delta K^{\mu\nu} (g\,;\,\zeta)- 2 \zeta^{[\mu}\Theta^{\nu]}(g;\, \delta g) + \sqrt{-g}{\bf A}^{\mu\nu}(g\,;\, \Lie_{\zeta}g, \delta g) \,,  \label{AsymtoticRel} \eeq
which is  identical with the off-shell ADT potential $Q_{ADT}^{\m\n}$ except for the additional term ${\bf A}^{\mu\nu}$.  

A couple of comments are in order.
\begin{itemize}
\item The generalized off-shell ADT current ${\bf J}^{\mu}_{ADT}$ depends only on the EOM, while the Noether current, $J^{\mu}$ depends on the Lagrangian and the symplectic current, $\omega^{\mu}$ does on the surface term $\Theta^{\mu}$. 
\item When $\zeta$ is an exact Killing vector,  the corresponding generalized current ${\bf J}^{\mu}_{ADT}$ reduces to the original off-shell ADT current $\CJ^{\mu}_{ADT}$.
\item When the background metric and the linearized metric $\delta g$ satisfy EOM and the linearized EOM, respectively, the generalized current ${\bf J}^{\mu}_{ADT}$ reduces to a symplectic current $\CJ^{\mu}_{\Delta}$. 
\item Though the above ${\bf Q}^{\mu\nu}_{ADT}$ is ambiguous up to the total derivative term $\p_{\rho} U^{\mu\nu\rho}$ by construction,  it is  irrelevant in our discussion on the asymptotic symmetry generators. 
\end{itemize}

%as $2\sqrt{-g}\CJ^{\mu}_{\Delta}\,;$
%
%\begin{equation} \label{}
%\delta (\sqrt{-g}\CE^{\mu\nu})~\Lie_{\zeta}g_{\mu\nu} - \Lie_{\zeta} (\sqrt{-g}\CE^{\mu\nu})~\delta g_{\mu\nu} =  - \p_{\mu}\omega^{\mu}(\Lie_{\zeta}g, \delta g)\,.
%\end{equation}
%

Quasi-local charge for an asymptotic Killing vector $\zeta$ may be defined  just like Eq.~(\ref{ADT}). Its infinitesimal variation  under an asymptotic Killing vector $\eta$  is given by
\bea  
\delta_{\eta} Q(\zeta) &\equiv &\frac{1}{16\pi G}\int dx_{\mu\nu} \sqrt{-g}\, {\bf Q}^{\mu\nu}_{ADT}(g\,;\, \Lie_\zeta g, \delta_{\eta}g) \\
%&=&\frac{1}{16\pi G}\int dx_{\mu\nu} \half \Big(\delta_{\eta}K^{\mu\nu}(\zeta) -2\zeta^{[\mu} \Theta^{\nu]}(\delta_\eta g) + {\bf A}^{\mu\nu}(g\,;\, \Lie_{\zeta}g,\delta g) \Big)  \nn \\ 
&=&\frac{1}{16\pi G}\int dx_{\mu\nu} \Big( \zeta^{[\mu} J^{\nu]}(\eta) -  \eta^{[\mu} J^{\nu]}(\zeta)  - \zeta^{[\mu}\eta^{\nu]}\sqrt{-g}\CL -\sqrt{-g}\zeta^{[\mu}\CE^{\nu]\rho}\eta_{\rho} \nn \\
&&\qquad \qquad \qquad \quad +\half  \sqrt{-g}{\bf A}^{\mu\nu} +\cdots \Big)\,, \nn
\eea
where $\,\cdots\,$ denotes irrelevant total derivative terms. In the last equality, we have used   the off-shell Noether current $J^{\mu}$ in Eq. (\ref{OffNoether}) along with the relation
\begin{equation} \label{}
\Lie_{\eta}K^{\mu\nu} (\zeta)= -2\eta^{[\mu}J^{\nu]}(\zeta) +3\p_{\alpha}\Big(\eta^{[\alpha}K^{\mu\nu]}(\zeta)\Big)\,.
\end{equation}
The asymptotic symmetry algebra can be constructed by using the above variational form of quasi-local charges. Since the detailed steps for this construction is completely parallel to those given, for instance, in~\cite{Barnich:2001jy,Compere:2007az,Compere:2008cv}, we will omit those.

%%%%%%%%%%%%%%%%%%%%%%%%%%%%%%%%%%%%%%%%%%%%%%
\subsection{Example: higher curvature gravity}
%%%%%%%%%%%%%%%%%%%%%%%%%%%%%%%%%%%%%%%%%%%%%%
In this section, we consider the higher curvature gravity as specific examples to apply our formulation of the generalized quasi-local charges.
In the higher curvature gravity $\CL = \CL(R, R^2, R_{\mu\nu}R^{\mu\nu}, \cdots)$, it is very useful to regard the Riemann tensor as an independent variable and to introduce 
\[
   P^{\mu\nu\rho\sigma} \equiv \frac{\p \CL}{\p R_{\mu\nu\rho\sigma}}\,.
\]
Then, one can show that the EOM expression $\CE^{\mu\nu}$, the off-shell Noether potential $K^{\mu\nu}$ and the surface term $\Theta^{\mu}$ take the  canonical forms of
\bea  \label{highercurv}
\CE^{\mu\nu} &=& P^{(\mu}_{~~ \alpha\beta\gamma}R^{\nu)\alpha\beta\gamma} - 2\nabla_{\rho}\nabla_{\sigma}P^{\rho(\mu\nu)\sigma} - \frac{1}{2}g^{\mu\nu}\CL\,, \\
K^{\mu\nu}(\zeta)&=& 2\sqrt{-g}\Big(P^{\mu\nu\rho\sigma}\nabla_{\rho}\zeta_{\sigma} - 2\zeta_{\sigma}\nabla_{\rho}P^{\mu\nu\rho\sigma}\Big)\,,   \nn \\
\Theta^{\mu}(\delta g) &=& 2\sqrt{-g}\Big(P^{\mu(\alpha\beta)\gamma}\nabla_{\gamma}\delta g_{\alpha\beta} - \delta g_{\alpha\beta}\nabla_{\gamma}P^{\mu(\alpha\beta)\gamma}\Big)\,. \nn 
\eea
Now, it is sufficient to obtain ${\bf A}^{\mu\nu}$ in order to find the quasi-local charge for an asymptotic Killing vector. 
To use our procedure in obtaining ${\bf A}^{\mu\nu}$ given in Eq.~(\ref{ATheta}), one may note that, after the repeated integration by parts, the antisymmetric part ${\bf A}^{\mu\nu}$ comes only from the variation of covariant derivatives in $\Theta^{\mu}(\Lie_{\xi}g)$. Specifically, the relevant part for the first term of $\Theta^{\mu}(\Lie_{\xi}g)$ in  (\ref{highercurv}) comes from the metric variation of the covariant derivative and is  given by
\begin{eqnarray}  
2P^{\mu(\alpha\beta)\nu}\delta \nabla_{\nu} \Lie_{\zeta} g_{\alpha\beta} &=& -4P^{\mu(\alpha\beta)\nu}~ \delta \Gamma^{\rho}_{\nu(\alpha}\, \Lie_{\zeta}g_{\beta)\rho} +  \cdots \nn \\
 &=&- 2\nabla_{\nu} \Big(P^{\mu(\nu\beta)\alpha}g^{\rho\sigma}\delta g_{\sigma\alpha} \Lie_{\zeta}g_{\rho\beta} - P^{\mu(\alpha\beta)\sigma}g^{\rho\nu}\delta g_{\sigma\alpha} \Lie_{\zeta}g_{\rho\beta} \Big)+ \cdots  \nn \\
&=& \nabla_{\nu}\Big[ -\frac{3}{4}P^{\mu\nu\alpha\rho}\,g^{\beta\sigma} (\Lie_{\zeta} g_{\alpha\beta}\, \delta g_{\rho\sigma} - \delta g_{\alpha\beta}\, \Lie_{\zeta} g_{\rho\sigma}) + g^{\rho[\mu} P^{\nu](\alpha\beta)\sigma}\delta g_{\alpha\beta}\, \Lie_{\zeta} g_{\rho\sigma} \Big] \nn \\
&& ~~ + \cdots \,,  \nn 
\end{eqnarray}
where $\cdots$ denotes the irrelevant terms, which are either symmetric part over $\mu\nu$-indices, ${\bf S}^{\mu\nu}$,  or non total derivative part, $\CV^{\alpha\beta\, |\, \rho}$. Just like the first term, 
the second term of $\Theta^{\mu}(\Lie_{\xi}g)$ in  (\ref{highercurv}) leads to the relevant terms as
\begin{eqnarray}  
2\Lie_{\zeta} g_{\alpha\beta}\, \delta \nabla_{\gamma}P^{\mu(\alpha\beta)\gamma} =  \nabla_{\nu}\Big[ \frac{3}{4}P^{\mu\nu\alpha\rho}\,g^{\beta\sigma} (\Lie_{\zeta} g_{\alpha\beta}\, \delta g_{\rho\sigma} - \delta g_{\alpha\beta}\, \Lie_{\zeta} g_{\rho\sigma})   + g^{\rho[\mu} P^{\nu](\alpha\beta)\sigma}\Lie_{\zeta} g_{\alpha\beta}\, \delta g_{\rho\sigma} \Big]  + \cdots\,.\nn 
\end{eqnarray}
Combining the above results, one can show that
\begin{equation} \label{}
{\bf A}^{\mu\nu}(\Lie_{\zeta}g, \delta g)  =  -\Big(\frac{3}{2}P^{\mu\nu\alpha\rho}\,g^{\beta\sigma}  + 2 g^{\rho[\mu} P^{\nu](\alpha\beta)\sigma}\Big)   (\Lie_{\zeta} g_{\alpha\beta}\, \delta g_{\rho\sigma} - \delta g_{\alpha\beta}\, \Lie_{\zeta} g_{\rho\sigma}) \,.
\end{equation}
It is very interesting to notice the complete consistency with the result  given in~\cite{Azeyanagi:2009wf}. This shows us that our construction in the case of the  higher curvature  theory of gravity is identical with the mathematical construction through the, so-called, horizontal homotopy operator in the BBC formalism. This equivalence  also tells us that the central charge in the asymptotic symmetry algebra should be the same as the one in the BBC formalism. It may be straightforward to check that our construction leads to the same results with the BBC formalism even in any higher derivative theory of gravity.

Just for concreteness, we present some detailed expressions in Einstein gravity. In this case, $P$-tensor is given by
\begin{equation} \label{}
P^{\mu\nu\rho\sigma}_{R} = g^{\rho[\mu}g^{\nu]\sigma}\,,
\end{equation}
and it turns out that  
\[   
{\bf A}^{\mu\nu}_{R}(\Lie_{\zeta}g, \delta g) =-\Big(g^{\mu(\alpha}g^{\beta)(\rho} g^{\sigma)\nu} - g^{\nu(\alpha}g^{\beta)(\rho}g^{\sigma)\mu} \Big)  \Big(\Lie_{\zeta}g_{\alpha\beta}\delta g_{\rho\sigma}  - \delta g_{\alpha\beta}\Lie_{\zeta} g_{\rho\sigma} \Big)\,.
\]
For any diffeomorphism paramter $\zeta$ ($h_{\mu\nu}=\delta g_{\mu\nu}$) in Einstein gravity
\beq \label{}
\sqrt{-g}Q^{\m\n}_R\equiv
\frac{1}{2}\delta K^{\mu\nu}_{R} - \zeta^{[\mu} \Theta^{\nu]}_{R} = \sqrt{-g}\Big(  \frac{1}{2}h \nabla^{[\mu}\zeta^{\nu]}    -\zeta^{[\mu} \nabla_{\alpha} h^{\nu]\alpha} + \zeta_{\alpha}\nabla^{[\mu}h^{\nu]\alpha}  + \zeta^{[\mu}\nabla^{\nu]} h  - h^{\alpha [\mu} \nabla_{\alpha} \zeta^{\nu]}  \Big)\,,
\eeq
and so, the generalized off-shell ADT potential for an asymptotic Killing vector $\zeta$ in Einstein gravity is given by
\bea
{\bf Q}^{\mu\nu}_R  &=&     Q^{\m\n}_R    + \frac{1}{2}{\bf A}^{\mu\nu}_{R} \nn \\
 &=&  \frac{1}{2}h \nabla^{[\mu}\zeta^{\nu]}    -\zeta^{[\mu} \nabla_{\alpha} h^{\nu]\alpha} + \zeta_{\alpha}\nabla^{[\mu}h^{\nu]\alpha}  + \zeta^{[\mu}\nabla^{\nu]} h   - \frac{1}{2} h^{\alpha [\mu} \nabla_{\alpha} \zeta^{\nu]} + \frac{1}{2}h^{\alpha [\mu} \nabla^{\nu]} \zeta_{\alpha}  \,.
\eea 
%
%where we have used  $E^{\mu\nu}_{R} = \sqrt{-g}{ \bf E}^{\mu\nu}_{R}$.

%%%%%%%%%%%%%%%%%%%%%%%%%
\section{Conclusion}
%%%%%%%%%%%%%%%%%%%%%%%%%
We have constructed the generalized off-shell ADT current and potential by using only the EOM and the linearized EOM expression. By connecting this construction to the covariant phase space method, we have presented a definite way to obtain the off-shell ADT current and potential from the given Lagrangian. Our construction can be applied to a generic higher derivative theory of gravity even with gravitational Chern-Simons terms. We have also indicated the relation of our construction to the BBC formalism. As a specific example, we have presented the generalized off-shell ADT potential explicitly for a higher curvature theory of gravity. Our construction shows us the usefulness of the off-shell or quasi-local formulation of conserved charges.  

In view of the generic structure of the generalized off-shell ADT current, our construction corresponds to the choice of  %By imposing identical vanishing condition of the above current and by using the fact that $\nabla_{(\nu_1}\cdots \nabla_{\nu_k} \Lie_{\zeta}g_{\alpha\beta)}$ as independent variables, one can see that
%
%\beq \label{}
 %\CM^{(\nu_1\nu_2\cdots\nu_{n-1}\alpha\beta)} + \nabla_{\mu}\CM^{\mu (\nu_1\cdots\nu_k\alpha\beta)}= \CM^{(\mu \nu_1\cdots \nu_n \alpha\beta)} =0 \,.
%\eeq
%
%These imply that
%%
%
%\begin{equation} \label{}
%\CM^{(\mu\alpha\beta)} = \CM^{(\mu \nu\alpha\beta)}= \CM^{\mu (\nu_1\cdots \nu_n \alpha\beta)}=0\,.
%\end{equation}
%
%Every lower moment $\CM$ in $\tilde{\CJ}$ has vanishing symmetric part except the highest one.
a  further symmetric form as follows
\begin{equation} \label{}
\CM^{\mu \nu_1\cdots \nu_k \alpha\beta} =   \CM^{(\mu \nu_1\cdots \nu_k) \alpha\beta}\,, \qquad k=1,2,\cdots \,.\end{equation}
This fact can be inferred from Eq.~(\ref{Criterion}) by noting that we can perform the iterative integration by parts  on $\Lie_{\zeta} g$ instead of $\delta g$.  By performing the integration by parts  further on the term ${\bf S}^{\rho\rho'}$, one can show that  $\CM^{\mu \nu_1\cdots \nu_n \alpha\beta}$ takes the  above form. In this sense, our construction may be regarded as the  most symmetric one.

As was shown in the higher curvature theory of gravity, our final results are equivalent to those from the BBC formalism.  To compare our construction to the BBC formalism, it is useful to recall that the BBC construction of the asymptotic symmetry generators starts from the so-called on-shell vanishing current $S^{\mu}(\zeta) = \sqrt{-g}\CE^{\mu\nu}\zeta_{\nu}$. One may note that our construction of the generalized off-shell ADT potential ${\bf Q}^{\mu\nu}_{ADT}$ given in Eqs.~(\ref{genoffADT})  and (\ref{genoffADTpot}) can be rewritten as  
\begin{equation} \label{}
\delta \Big(2\sqrt{-g}\CE^{\mu\nu}\zeta_{\nu}\Big) = -2\sqrt{-g}\CJ^{\mu}_{\Delta}(\Lie_{\zeta}g,\, \delta g) +  \sqrt{-g}\zeta^{\mu}\CE^{\alpha\beta}\delta g_{\alpha\beta} + \p_{\nu}\Big(2\sqrt{-g}{\bf Q}^{\mu\nu}_{ADT}\Big)\,. 
\end{equation}
This  shows that our generalized off-shell ADT potential ${\bf Q}^{\mu\nu}_{ADT}$ corresponds to  the potential $k^{\mu\nu}$ in the BBC formalism~\cite{Barnich:2001jy,Compere:2007az}.
In conjunction with the equivalence in the case of a higher curvature theory of gravity,  this form strongly indicates the formal equivalence between our construction of the generalized off-shell ADT potential and the  BBC formalism, though it does not prove the equivalence.   On the other hand, there are some differences between two constructions. In our construction we have not used an on-shell condition on the metric $g$ and  do not need any canonical choices of the surface term $\Theta^{\mu}$ or the off-shell Noether potential $K^{\mu\nu}$. Any ambiguity in such terms should be canceled in Eq.~(\ref{AsymtoticRel}) by construction. In the BBC formalism, a priori ambiguous quantities like the  $\Theta$-term are fixed by the horizontal homotopy operator while we do not need such canonical choice. However,  it seems very plausible to expect the formal equivalence between them in consideration of the final expression for the potentials. It would be very interesting to prove their equivalence. 

As a further direction, it would be also very interesting to study physics in~\cite{Strominger:2013jfa,He:2014laa,Barnich:2013axa} by using our construction.

%Let us introduce $\CW^{\mu}$ as follows
%
%\begin{equation} \label{}
%\delta (\sqrt{-g}\CE^{\mu\nu})~\Lie_{\zeta}g_{\mu\nu} = \CF^{\alpha\beta}(\zeta)\,\delta g_{\alpha\beta}  -\p_{\mu}\CW^{\mu}(\Lie_{\zeta}g,\, \delta g)\,,
%\end{equation}
%
%where $\CW$ is obtained as follows. 

\vskip 1cm
\centerline{\large \bf Acknowledgments}
\vskip0.5cm

{SH was supported by the National Research Foundation of Korea(NRF) grant funded 
by the Korea government(MEST) with the grant number  2012046278 and the grant number 2013-110892.  S.-H.Yi was supported by the National Research Foundation of Korea(NRF) grant funded by the Korea government(MOE) (No.  2012R1A1A2004410).}
\newpage

%%%%%%%%%%%%%%%%%%%%%%%%%
\section*{Appendix A}
%%%%%%%%%%%%%%%%%%%%%%%%%
%%%%%%%%%%%%%%%%%%%%%%%%%%%%%%
\renewcommand{\theequation}{A.\arabic{equation}}
  \setcounter{equation}{0}
%%%%%%%%%%%%%%%%%%%%%%%%%%%%%%%%%%%%%%%%%%%%%%%%%%%
Let us take another generic variation of the expression (\ref{firstvary}) as
\begin{equation} \label{}
\delta_2\delta_1 I[g] = \frac{1}{16\pi G}\int d^Dx \Big[-\delta_2(\sqrt{-g}\CE^{\mu\nu})\delta_1 g_{\mu\nu} - \sqrt{-g}\CE^{\mu\nu} \delta_2\delta_1 g_{\mu\nu} + \p_{\mu}\Big(\delta_2\Theta^{\mu}(\delta_1 g)\Big) \Big]\,.
\end{equation}
Through the extension of the relation~(\ref{Intbypart}) to the case of the generic variation, one can set the above double variation in the form of 
\begin{eqnarray} \label{}
\delta_2\delta_1 I[g] &=&\frac{1}{16\pi G} \int d^Dx \Big[ -\sqrt{-g}\CF^{\mu\nu}(\delta_1 g)~ \delta_2 g_{\mu\nu} - \sqrt{-g}\CE^{\mu\nu} \delta_2\delta_1 g_{\mu\nu} \nn \\
&&\qquad \qquad \qquad \qquad \qquad + \, \p_{\mu}\Big(\delta_2\Theta^{\mu}(\delta_1 g) - \sqrt{-g}\CH^{\mu}(\delta_1g, \delta_2 g) \Big) ~\Big] \nn \\
&=&\frac{1}{16\pi G} \int d^Dx \Big[-\delta_1(\sqrt{-g}\CE^{\mu\nu})\delta_2 g_{\mu\nu} - \sqrt{-g}\CE^{\mu\nu} \delta_1\delta_2 g_{\mu\nu} + \p_{\mu}\Big(\delta_1\Theta^{\mu}(\delta_2 g)\Big) ~ \Big]\,, \nn
\end{eqnarray}
where the second equality comes from the commuting relation between two generic variations as $\delta_1\delta_2 I[g] = \delta_2\delta_1 I[g]$.
Now, let us take the variation $\delta_2 g$ to be generic but  compactly supported only in the bulk. In other words, $\delta_2 g$ is taken to be decaying sufficiently fast at the boundary of the region of interest.  This choice tells us that we can ignore the surface term for such a variation  $\delta_2 g$. Under this condition with the relation  $\delta_1\delta_2 g_{\mu\nu} = \delta_2\delta_1 g_{\mu\nu}$, we can obtain the relation
\begin{equation} \label{}
\sqrt{-g} \CF^{\mu\nu}(\delta g) = \delta (\sqrt{-g}\CE^{\mu\nu})\,,
\end{equation}
which should hold for an arbitrary metric variation $\delta g_{\mu\nu}$.
%Note that an asymptotic Killing vector $\Lie_{\xi}g_{\mu\nu}$ satisfies the above fast decaying condition and so the relation~(\ref{Ida}) is proved. 
By using the explicit form of $\CF^{\mu\nu}$ given in Eq.~(\ref{Fexp}) in conjunction with Eq.~(\ref{deltaE}), one can obtain the following identity
\begin{equation} \label{Idb}
  f^{\alpha\beta\mu\nu\,|\, \rho_1\cdots \rho_{\ell}} =\sum_{k=\ell}^{n}(-1)^{k}{k \choose \ell}\nabla_{\rho_{\ell+1}}\cdots \nabla_{\rho_k} f^{\mu\nu\alpha\beta\,|\, \rho_1\cdots \rho_k}\,.
\end{equation}
In the case of  Einstein gravity, this identity implies that the non-vanishing terms, $f^{\mu\nu\alpha\beta\,|\, \rho_1\rho_2}$ satisfy 
\begin{equation} \label{}
f^{\mu\nu\alpha\beta\,|\, \rho_1\rho_2} = f^{\alpha\beta\mu\nu\,|\, \rho_1\rho_2}
\end{equation}
%

%%%%%%%%%%%%%%%%%%%%%%%%%
\section*{Appendix B}
%%%%%%%%%%%%%%%%%%%%%%%%%
%%%%%%%%%%%%%%%%%%%%%%%%%%%%%%
\renewcommand{\theequation}{B.\arabic{equation}}
  \setcounter{equation}{0}
%%%%%%%%%%%%%%%%%%%%%%%%%%%%%%%%%%%%%%%%%%%%%%%%%%%
In this appendix we show the main steps leading  Eq. (\ref{CJexpF}). 
For our convenience, we may represent  the  expressions of $\CF^{\a\b}$ and $\CJ_{\Delta}^{\r}$  in Eqs. (\ref{ours}) and (\ref{sang}) compactly as  
\bea 
\CF^{\alpha\beta} &=& \sum_{k=0}^{n}(-1)^k\nabla_{\rho_1}\cdots\nabla_{\rho_k}( f^{\mu\nu\alpha\beta\,|\, \rho_1\rho_2\cdots\rho_k}\, \Lie_{\zeta}g_{\mu\nu})\,,  \nn \\
\CJ^{\rho}_{\Delta}(\Lie_{\zeta}g,\, \delta g) &=& \frac{1}{2}\sum_{k=1}^{n}\sum^{k}_{l=1}(-1)^{\ell}\nabla_{\rho_l}\cdots\nabla_{\rho_2}( f^{\mu\nu\alpha\beta \,|\, \rho \rho_2\cdots \rho_{k}}\, \Lie_{\zeta}g_{\mu\nu})~ \nabla_{\rho_{l+1}}\cdots \nabla_{\rho_k}\delta g_{\alpha\beta}  \,. \nn
\eea
By using the binomial expansion 
\[   
\nabla_{(\rho_1}\cdots \nabla_{\rho_k)}(AB) = \sum_{p=0}^{k}{ k \choose p}\nabla_{(\rho_1}\cdots \nabla_{\rho_{p}}A~ \nabla_{\rho_{p+1}}\cdots \nabla_{\rho_k)}B\,,
\]
one can rewrite those as
\begin{eqnarray}  
\CF^{\alpha\beta} &=& \sum_{k=0}^{n}\sum_{p=0}^{k}(-1)^k{ k \choose p}\nabla_{\rho_{p+1}}\cdots \nabla_{\rho_k}f^{\mu\nu\alpha\beta\,|\,\rho_1\cdots \rho_k}~  \nabla_{\rho_1}\cdots \nabla_{\rho_{p}}\Lie_{\zeta}g_{\mu\nu}  \nn  \\
\CJ^{\rho}_{\Delta}(\Lie_{\zeta}g,\, \delta g) &=& \frac{1}{2}  \sum_{k=1}^{n}\sum^{k}_{\ell=1}\sum_{p=0}^{l-1}(-1)^{\ell}{ \ell-1 \choose p} \nabla_{\rho_{p+2}} \cdots \nabla_{\rho_l} f^{\mu\nu\alpha\beta \,|\, \rho \rho_2\cdots \rho_{k} }~ \nn \\
&& \qquad \qquad  \qquad  \qquad \qquad  \qquad  \qquad  \times  \nabla_{\rho_2}\cdots\nabla_{\rho_{p+1}}\Lie_{\zeta}g_{\mu\nu} ~  \nabla_{\rho_{l+1}}\cdots \nabla_{\rho_k}\delta g_{\alpha\beta}\,.  \nn 
\end{eqnarray}
After rearranging the order of summation, one can obtain more useful expression as
\begin{eqnarray}  
\CF^{\alpha\beta}
&=&  \sum_{p=0}^{n}\sum_{k=p}^{n}(-1)^k{ k \choose p}\nabla_{\rho_{p+1}}\cdots \nabla_{\rho_k} f^{\mu\nu\alpha\beta\,|\, \rho_1\cdots \rho_k}~ \nabla_{\rho_1}\cdots \nabla_{\rho_{p}}\Lie_{\zeta}g_{\mu\nu}\,,   \label{Fexpa} \\
\CJ^{\rho}_{\Delta} &=& \frac{1}{2} \sum_{p=1}^{n}\sum_{k=p}^{n}\sum_{q=0}^{p-1}(-1)^{p+k+q+1} {k-p +q \choose q}   \nabla_{\rho_{p+1}}\cdots \nabla_{\rho_k} f^{\mu\nu\alpha\beta\,|\, \rho\rho_2\cdots\rho_k}     \label{Wexp}\\ 
&& \qquad \qquad \qquad \qquad \qquad  \qquad \qquad \times  \nabla_{\rho_{p-q+1}}\cdots\nabla_{\rho_p}\Lie_{\zeta}g_{\mu\nu}\, \nabla_{\rho_2}\cdots \nabla_{\rho_{p-q}}\delta g_{\alpha\beta}\,. \nn 
\end{eqnarray}

Now we would like to show the anti-symmetric property of $\CJ^{\mu}_{\Delta}(\Lie_{\zeta}g,\, \delta g) $ over its arguments. To this purpose,  we need differential relations  among $f$ functions.
Firstly, one may note that

%%%%%%
%%%%%%
%
%
\begin{eqnarray}  
&& \sum_{\ell=p}^{n}(-1)^{\ell} {\ell - p  + q \choose q} \nabla_{\rho_{p+1}}\cdots \nabla_{\rho_\ell} f^{\alpha\beta\mu\nu\,|\, \rho_1\cdots \rho_\ell}   \\
&=& \sum_{\ell =p}^{n}\sum_{k=\ell }^{n}(-1)^{\ell+k} {\ell - p  + q \choose q} {k\choose \ell}\nabla_{\rho_{p+1}}\cdots \nabla_{\rho_k} f^{\mu\nu\alpha\beta\,|\, \rho_1\cdots \rho_k}\,.  \nn 
\\
&=&  \sum_{k=p}^{n}(-1)^k\bigg[\sum_{m=p}^{k}(-1)^{m}{k\choose m} {m - p +q  \choose q}\bigg] \nabla_{\rho_{p+1}}\cdots \nabla_{\rho_k} f^{\mu\nu\alpha\beta\,|\, \rho_1\cdots \rho_k}\,,    \nn 
\end{eqnarray}
where we have used the identities~(\ref{Idb}) in the first equality and  rearranged the order of the summations in the second equality.  
Secondly,  by using  the  binomial identity 
\begin{equation} \label{}
\sum_{m=p}^{k}(-1)^m{k\choose m} {m - p +q  \choose q} = (-1)^p { k - q-1 \choose p-q-1}\,,
\end{equation}
we obtain the identity
\begin{eqnarray}   \label{Idc}
&& \sum_{\ell=p}^{n}(-1)^{\ell} {\ell - p  + q \choose q} \nabla_{\rho_{p+1}}\cdots \nabla_{\rho_\ell} f^{\alpha\beta\mu\nu\,|\, \rho_1\cdots \rho_\ell}   \\ && =~  \sum_{k=p}^{n}(-1)^{k+p} { k - q -1  \choose p- q-1} \nabla_{\rho_{p+1}}\cdots \nabla_{\rho_k} f^{\mu\nu\alpha\beta\,|\, \rho_1\cdots \rho_k}\,.  \nn
\end{eqnarray}

By using the above identity into  Eq.~(\ref{Wexp}), one can find the manifestly antisymmetric form of  $\CJ^{\r}_{\Delta}$ over its arguments as
\begin{eqnarray}  
\CJ^{\rho}_{\Delta}(\Lie_{\zeta}g, \delta g) &=&\frac{1}{2} \sum_{p=1}^{n}\sum_{k=p}^{n}\sum_{q=0}^{\big[\frac{p-1}{2}\big]}(-1)^{p+k+q+1} {k-p +q \choose q}~   \nabla_{\rho_{p+1}}\cdots \nabla_{\rho_k} f^{\mu\nu\alpha\beta\,|\, \rho\rho_2\cdots\rho_k}     \label{CJexpF}\\ 
&&  \quad   \times \Big(1- \frac{1}{2}\,\delta_{2q+1, p}\Big)  \Big(\nabla_{\rho_{p-q+1}}\cdots\nabla_{\rho_p}\Lie_{\zeta}g_{\mu\nu}\, \nabla_{\rho_2}\cdots \nabla_{\rho_{p-q}}\delta g_{\alpha\beta}  \nn \\
&& \qquad \qquad \qquad \qquad \qquad \qquad  -~ \nabla_{\rho_2}\cdots \nabla_{\rho_{p-q}}\Lie_{\zeta} g_{\alpha\beta}  \nabla_{\rho_{p-q+1}}\cdots\nabla_{\rho_p}\delta g_{\mu\nu}\, \Big)\,. \nn 
\end{eqnarray}
%%

%%%%%%%%%%%%%%%%%%%%%%%%%%%%%%%%%%%%%%%%%%%%%%%%
%%%%%%%%%%%%%%%             References         %%%%%%%%%%%%%%%%
%%%%%%%%%%%%%%%%%%%%%%%%%%%%%%%%%%%%%%%%%%%%%%%%
%%%%%%%%%%%%%%%%%%%%%%%%%%%%%%%%%%%%%%%%%%%%%

%%%%%%%%%%%%%%%%%%%%%%%%%%%%%%%%%%%%%%%%%%

\begin{thebibliography}{99} 
%%%%%%%%%%%%%%%%%%%%%%%%%%%%%%%%%%%%%%%%%%%%%%\cite{Maldacena:1997re}
\bibitem{Maldacena:1997re} 
  J.~M.~Maldacena,
  ``The Large N limit of superconformal field theories and supergravity,''
  Adv.\ Theor.\ Math.\ Phys.\  {\bf 2}, 231 (1998)
  [hep-th/9711200].
  %%CITATION = HEP-TH/9711200;%%
  %9645 citations counted in INSPIRE as of 07 Mar 2014


%\cite{Brown:1986nw}
\bibitem{Brown:1986nw} 
  J.~D.~Brown and M.~Henneaux,
  ``Central Charges in the Canonical Realization of Asymptotic Symmetries: An Example from Three-Dimensional Gravity,''
  Commun.\ Math.\ Phys.\  {\bf 104}, 207 (1986).
  %%CITATION = CMPHA,104,207;%%
  %991 citations counted in INSPIRE as of 07 Mar 2014


%\cite{Cardy:1986ie}
\bibitem{Cardy:1986ie} 
  J.~L.~Cardy,
  ``Operator Content of Two-Dimensional Conformally Invariant Theories,''
  Nucl.\ Phys.\ B {\bf 270}, 186 (1986).
  %%CITATION = NUPHA,B270,186;%%
  %884 citations counted in INSPIRE as of 07 Mar 2014


%\cite{Strominger:1996sh}
\bibitem{Strominger:1996sh} 
  A.~Strominger and C.~Vafa,
  ``Microscopic origin of the Bekenstein-Hawking entropy,''
  Phys.\ Lett.\ B {\bf 379}, 99 (1996)
  [hep-th/9601029].
  %%CITATION = HEP-TH/9601029;%%
  %1792 citations counted in INSPIRE as of 07 Mar 2014


%\cite{Guica:2008mu}
\bibitem{Guica:2008mu} 
  M.~Guica, T.~Hartman, W.~Song and A.~Strominger,
  ``The Kerr/CFT Correspondence,''
  Phys.\ Rev.\ D {\bf 80}, 124008 (2009)
  [arXiv:0809.4266 [hep-th]].
  %%CITATION = ARXIV:0809.4266;%%
  %313 citations counted in INSPIRE as of 07 Mar 2014


%\cite{Szabados:2004vb}
\bibitem{Szabados:2004vb} 
  L.~B.~Szabados,
  ``Quasi-Local Energy-Momentum and Angular Momentum in GR: A Review Article,''
  Living Rev.\ Rel.\  {\bf 7}, 4 (2004).
  %%CITATION = 00222,7,4;%%
  %120 citations counted in INSPIRE as of 07 Mar 2014

%\cite{Hollands:2005wt}
\bibitem{Hollands:2005wt} 
  S.~Hollands, A.~Ishibashi and D.~Marolf,
  ``Comparison between various notions of conserved charges in asymptotically AdS-spacetimes,''
  Class.\ Quant.\ Grav.\  {\bf 22}, 2881 (2005)
  [hep-th/0503045].
  %%CITATION = HEP-TH/0503045;%%
  %81 citations counted in INSPIRE as of 07 Mar 2014

%\cite{Abbott:1981ff}
\bibitem{Abbott:1981ff} 
  L.~F.~Abbott and S.~Deser,
  ``Stability of Gravity with a Cosmological Constant,''
  Nucl.\ Phys.\ B {\bf 195}, 76 (1982).
  %%CITATION = NUPHA,B195,76;%%
  %564 citations counted in INSPIRE as of 07 Mar 2014


%\cite{Abbott:1982jh}
\bibitem{Abbott:1982jh} 
  L.~F.~Abbott and S.~Deser,
  ``Charge Definition in Nonabelian Gauge Theories,''
  Phys.\ Lett.\ B {\bf 116}, 259 (1982).
  %%CITATION = PHLTA,B116,259;%%
  %78 citations counted in INSPIRE as of 07 Mar 2014


%\cite{Deser:2002rt}
\bibitem{Deser:2002rt} 
  S.~Deser and B.~Tekin,
  ``Gravitational energy in quadratic curvature gravities,''
  Phys.\ Rev.\ Lett.\  {\bf 89}, 101101 (2002)
  [hep-th/0205318].
  %%CITATION = HEP-TH/0205318;%%
  %126 citations counted in INSPIRE as of 07 Mar 2014


%\cite{Deser:2002jk}
\bibitem{Deser:2002jk} 
  S.~Deser and B.~Tekin,
  ``Energy in generic higher curvature gravity theories,''
  Phys.\ Rev.\ D {\bf 67}, 084009 (2003)
  [hep-th/0212292].
  %%CITATION = HEP-TH/0212292;%%
  %187 citations counted in INSPIRE as of 07 Mar 2014


%\cite{Lee:1990nz}
\bibitem{Lee:1990nz} 
  J.~Lee and R.~M.~Wald,
  ``Local symmetries and constraints,''
  J.\ Math.\ Phys.\  {\bf 31}, 725 (1990).
  %%CITATION = JMAPA,31,725;%%
  %195 citations counted in INSPIRE as of 07 Mar 2014


%\cite{Wald:1993nt}
\bibitem{Wald:1993nt} 
  R.~M.~Wald,
  ``Black hole entropy is the Noether charge,''
  Phys.\ Rev.\ D {\bf 48}, 3427 (1993)
  [gr-qc/9307038].
  %%CITATION = GR-QC/9307038;%%
  %852 citations counted in INSPIRE as of 07 Mar 2014


%\cite{Iyer:1994ys}
\bibitem{Iyer:1994ys} 
  V.~Iyer and R.~M.~Wald,
  ``Some properties of Noether charge and a proposal for dynamical black hole entropy,''
  Phys.\ Rev.\ D {\bf 50}, 846 (1994)
  [gr-qc/9403028].
  %%CITATION = GR-QC/9403028;%%
  %723 citations counted in INSPIRE as of 07 Mar 2014


%\cite{Wald:1999wa}
\bibitem{Wald:1999wa} 
  R.~M.~Wald and A.~Zoupas,
  ``A General definition of 'conserved quantities' in general relativity and other theories of gravity,''
  Phys.\ Rev.\ D {\bf 61}, 084027 (2000)
  [gr-qc/9911095].
  %%CITATION = GR-QC/9911095;%%
  %112 citations counted in INSPIRE as of 07 Mar 2014

%\cite{Carlip:1999cy}
\bibitem{Carlip:1999cy} 
  S.~Carlip,
  ``Entropy from conformal field theory at Killing horizons,''
  Class.\ Quant.\ Grav.\  {\bf 16}, 3327 (1999)
  [gr-qc/9906126].
  %%CITATION = GR-QC/9906126;%%
  %197 citations counted in INSPIRE as of 12 Mar 2014
  
%\cite{Koga:2001vq}
\bibitem{Koga:2001vq} 
  J.~-i.~Koga,
  ``Asymptotic symmetries on Killing horizons,''
  Phys.\ Rev.\ D {\bf 64}, 124012 (2001)
  [gr-qc/0107096].
  %%CITATION = GR-QC/0107096;%%
  %42 citations counted in INSPIRE as of 07 Mar 2014


%\cite{Barnich:2001jy}
\bibitem{Barnich:2001jy} 
  G.~Barnich and F.~Brandt,
  ``Covariant theory of asymptotic symmetries, conservation laws and central charges,''
  Nucl.\ Phys.\ B {\bf 633}, 3 (2002)
  [hep-th/0111246].
  %%CITATION = HEP-TH/0111246;%%
  %193 citations counted in INSPIRE as of 07 Mar 2014


%\cite{Barnich:2007bf}
\bibitem{Barnich:2007bf} 
  G.~Barnich and G.~Compere,
  ``Surface charge algebra in gauge theories and thermodynamic integrability,''
  J.\ Math.\ Phys.\  {\bf 49}, 042901 (2008)
  [arXiv:0708.2378 [gr-qc]].
  %%CITATION = ARXIV:0708.2378;%%
  %98 citations counted in INSPIRE as of 07 Mar 2014


%\cite{Compere:2007az}
\bibitem{Compere:2007az} 
  G.~Compere,
  ``Symmetries and conservation laws in Lagrangian gauge theories with applications to the mechanics of black holes and to gravity in three dimensions,''
  arXiv:0708.3153 [hep-th].
  %%CITATION = ARXIV:0708.3153;%%
  %45 citations counted in INSPIRE as of 07 Mar 2014


%\cite{Hotta:2008yq}
\bibitem{Hotta:2008yq} 
  K.~Hotta, Y.~Hyakutake, T.~Kubota and H.~Tanida,
  ``Brown-Henneaux's Canonical Approach to Topologically Massive Gravity,''
  JHEP {\bf 0807}, 066 (2008)
  [arXiv:0805.2005 [hep-th]].
  %%CITATION = ARXIV:0805.2005;%%
  %47 citations counted in INSPIRE as of 07 Mar 2014


%\cite{Compere:2008cv}
\bibitem{Compere:2008cv} 
  G.~Compere and S.~Detournay,
  ``Semi-classical central charge in topologically massive gravity,''
  Class.\ Quant.\ Grav.\  {\bf 26}, 012001 (2009)
  [Erratum-ibid.\  {\bf 26}, 139801 (2009)]
  [arXiv:0808.1911 [hep-th]].
  %%CITATION = ARXIV:0808.1911;%%
  %71 citations counted in INSPIRE as of 07 Mar 2014


%\cite{Azeyanagi:2009wf}
\bibitem{Azeyanagi:2009wf} 
  T.~Azeyanagi, G.~Compere, N.~Ogawa, Y.~Tachikawa and S.~Terashima,
  ``Higher-Derivative Corrections to the Asymptotic Virasoro Symmetry of 4d Extremal Black Holes,''
  Prog.\ Theor.\ Phys.\  {\bf 122}, 355 (2009)
  [arXiv:0903.4176 [hep-th]].
  %%CITATION = ARXIV:0903.4176;%%
  %60 citations counted in INSPIRE as of 07 Mar 2014


%\cite{Kim:2013zha}
\bibitem{Kim:2013zha} 
  W.~Kim, S.~Kulkarni and S.~-H.~Yi,
  ``Quasi-Local Conserved Charges in Covariant Theory of Gravity,''
  Phys.\ Rev.\ Lett.\  {\bf 111}, 081101 (2013)
  [arXiv:1306.2138 [hep-th]].
  %%CITATION = ARXIV:1306.2138;%%
  %3 citations counted in INSPIRE as of 07 Mar 2014


%\cite{Kim:2013cor}
\bibitem{Kim:2013cor} 
  W.~Kim, S.~Kulkarni and S.~-H.~Yi,
  ``Quasilocal Conserved Charges with a Gravitational Chern-Simons Term,''
  Phys.\ Rev.\ D {\bf 88}, 124004 (2013)
  [arXiv:1310.1739 [hep-th]].
  %%CITATION = ARXIV:1310.1739;%%
  %1 citations counted in INSPIRE as of 07 Mar 2014


%\cite{Bouchareb:2007yx}
\bibitem{Bouchareb:2007yx} 
  A.~Bouchareb and G.~Clement,
  ``Black hole mass and angular momentum in topologically massive gravity,''
  Class.\ Quant.\ Grav.\  {\bf 24}, 5581 (2007)
  [arXiv:0706.0263 [gr-qc]].
  %%CITATION = ARXIV:0706.0263;%%
  %78 citations counted in INSPIRE as of 07 Mar 2014


%\cite{Nam:2010ub}
\bibitem{Nam:2010ub} 
  S.~Nam, J.~-D.~Park and S.~-H.~Yi,
  ``Mass and Angular momentum of Black Holes in New Massive Gravity,''
  Phys.\ Rev.\ D {\bf 82}, 124049 (2010)
  [arXiv:1009.1962 [hep-th]].
  %%CITATION = ARXIV:1009.1962;%%
  %20 citations counted in INSPIRE as of 07 Mar 2014


%\cite{Barnich:2003xg}
\bibitem{Barnich:2003xg} 
  G.~Barnich,
  ``Boundary charges in gauge theories: Using Stokes theorem in the bulk,''
  Class.\ Quant.\ Grav.\  {\bf 20}, 3685 (2003)
  [hep-th/0301039].
  %%CITATION = HEP-TH/0301039;%%
  %52 citations counted in INSPIRE as of 07 Mar 2014


%\cite{Brown:1992br}
\bibitem{Brown:1992br} 
  J.~D.~Brown and J.~W.~York, Jr.,
  ``Quasilocal energy and conserved charges derived from the gravitational action,''
  Phys.\ Rev.\ D {\bf 47}, 1407 (1993)
  [gr-qc/9209012].
  %%CITATION = GR-QC/9209012;%%
  %847 citations counted in INSPIRE as of 07 Mar 2014


%\cite{Strominger:2013jfa}
\bibitem{Strominger:2013jfa} 
  A.~Strominger,
  ``On BMS Invariance of Gravitational Scattering,''
  arXiv:1312.2229 [hep-th].
  %%CITATION = ARXIV:1312.2229;%%
  %4 citations counted in INSPIRE as of 07 Mar 2014


%\cite{He:2014laa}
\bibitem{He:2014laa} 
  T.~He, V.~Lysov, P.~Mitra and A.~Strominger,
  ``BMS supertranslations and Weinberg's soft graviton theorem,''
  arXiv:1401.7026 [hep-th].
  %%CITATION = ARXIV:1401.7026;%%
  %2 citations counted in INSPIRE as of 07 Mar 2014


%\cite{Barnich:2013axa}
\bibitem{Barnich:2013axa} 
  G.~Barnich and Céd.~Troessaert,
  ``Comments on holographic current algebras and asymptotically flat four dimensional spacetimes at null infinity,''
  JHEP {\bf 1311}, 003 (2013)
  [arXiv:1309.0794 [hep-th]].
  %%CITATION = ARXIV:1309.0794;%%
  %4 citations counted in INSPIRE as of 07 Mar 2014
  


%%%%%%%%%%%%%%%%%%%%%%%%%%%%%%%%%%%%%%%%%%%%%%%%%%%
\end{thebibliography}
\end{document}